\documentclass[
aps,%
12pt,%
final,%
notitlepage,%
oneside,%
onecolumn,%
nobibnotes,%
nofootinbib,%
superscriptaddress,%
noshowpacs,%
centertags]%
{revtex4}
\usepackage{graphicx}

\begin{document}
%\selectlanguage{russian}
%\end{document}

\title{GPDs and gravitational form factors of nucleons
(quark and gluon contributions)}

\author{\firstname{O.V.}~\surname{Selyugin}}
% ����� ࠧ������ �� ��ப� ������⢫����� ��⮬�����᪨ ��� �������� \\
\email{selugin@theor.jinr.ru}
\affiliation{
BLTP, JINR, Dubna, Russia
}%
\author{\firstname{O.V.}~\surname{Teryaev}}
\email{teryaev@theor.jinr.ru}
\affiliation{BLTP, JINR, Dubna, Russia
}%

%\author{\firstname{�.~�.}~\surname{���⨩-�����}}
%\email{Third.Author@univ.edu}
%%\noaffiliation % �᫨ � ������ ����� ࠡ��� �� 㪠�뢠����
%\affiliation{%
%����� ࠡ��� �/��� ����� ����쥣� ������
%}%

%\author{\firstname{�.~�.}~\surname{��⢥����-�����}}
%\email{Fourth.Author@inst.ras.ru}
%\affiliation{%
%����� ࠡ��� �/��� ����� ��⢥�⮣� ������
%}%

%\date{\today}
%\today ����⠥� c������譥� ��᫮

\begin{abstract}
 Taking into account the recent parameterizations  of parton distribution functions (PDFs), % obtained in the recent time,
 the momentum transfer dependence of generalized parton distributions (GPDs)
 of nucleons is obtained  in the limit $\xi \rightarrow 0$. The  gravitational  quark and gluon
  form factors of nucleons are calculated. It is shown that the gluon gravitational radius of the nucleon
  is comparable to the electromagnetic radius of the proton.
  The power dependence of form factors is investigated. As a result, it was obtained that
  the quark gravitational form factor is reproduced by the dipole form, while the form of
   the gluon gravitation form factor corresponds to the triple form.

\end{abstract}

\maketitle

\section{Introduction}
%\end{document} %Maybe .................... %\verb+\input maik.rty+ %\verb+\endinput+
   The structure of  hadrons reflected in generalized parton distributions
   is now one of the most interesting questions                   %{Burk-N1}
   of the physics of strong interactions (see, for example \cite{Burk-N1,DD-N2}.
   It is tightly connected with the spin physics of the hadron \cite{Ter-gpd-sp}.
   Some modern accelerator experiments have developed extensive programs  for deep studies of
   different issues related to this problem.  For example, the Jefferson Laboratory/Electron-Ion Collider (EIC) team uses Deeply Virtual Compton Scattering reactions to extract  generalized parton distributions \cite{NPNews32},  and
   future experiments at SPD of NICA-JINR \cite{NICA}.

     Many collaborations have already
  obtained  the form of  parton distribution functions (PDF) using  data obtained at HERA and LHC .
%  Besides this main point of the modern picture of the hadron structure, which
  If the old form for describing   hadron structure, parton distributions functions (PDF), depended
  only on the Bjorken longitudinal variable $x$,
   %there were introduced a number of  other
    more complicated functions,  generalized parton distributions  (GPDs),  depending on $ x$,
    momentum transfer $t$
  and the skewness parameter $\xi$ were introduced.

     The remarkable property of $GPDs(x,\xi,t)$ is that the integration of  different momenta of
     GPDs over $x$ gives us  different hadron form factors % \cite{Ji97,R97}. %
      \cite{Mul94,Ji97,R97}.
     The $x$ dependence of $GPDs$ is mostly determined by the standard
     parton distribution functions (PDFs),
     which were obtained by  various collaborations from the analysis of deep-inelastic processes.
% Many different forms
%  of the $t$-dependence of GPDs were proposed.
%   In the quark di-quark model \cite{Liuti1} the form of  GPDs
%   consists of three parts - PDFs, function distribution and the Regge-like function.
% In other works (see e.g. \cite{Kroll04}),
%  the description of the $t$-dependence of  GPDs  was developed
%  in a  more complicated picture using the polynomial forms with respect to $x$.
%
 Since the GPD is not known a priori, one seeks for models of GPDs
  %(see e.g. \cite{3})
  based on general constraints on their analytic and asymptotic
 behavior. The calculated scattering amplitudes (cross sections) are then compared with the data to confirm, modify or reject the chosen form of the GPDs.

  Commonly, the form of $GPDs(x,\xi,t)$ is determined  through the
  exclusive deep inelastic processes  of type $\gamma^*p\rightarrow Vp,$ where $V$ stands for a  photon
   or a vector meson. However,
  such processes have a narrow region of momentum transfer and
  in most models the $t$-dependence of GPDs is taken in  factorization form
  with the Gaussian form of the $t$-dependence. Really, this form of $GPDs(x,\xi,t)$ can not be used
  to build  space structure of the hadrons,
   as for that one needs to integrate over $t$ in a maximally wide region.
    %(if say exactly from zero to infinity).
   In the standard definition, it is connected to parton distributions of quarks and, especially, gluons.

\section{GPDs and hadron form factors}
  %  However,
  The hadron
 form factors are related to the $GPDs(x,\xi,t)$
 by the %standard
 sum rules \cite{Ji97}
 \begin{eqnarray}
 F_1^q(t)=\int_{-1}^1 dx H^q(x,\xi=0,t),  \ \ \ \ F_2^q(t)=\int_{-1}^1 dx E^q(x,\xi=0,t).
 \end{eqnarray}

 The integration region can be reduced to positive values of
 $x,~0<x<1$ by the following combination of non-forward parton
 densities %\cite{Rad1,GPRV}
  ${\cal H}^q(x,t)=H^q(x,0,t)+H^q(-x,0,t)$,
   ${\cal E}^q(x,t)=E^q(x,0,t)+E^q(-x,0,t)$
    providing
 $F^q_1(t)=\int_0^1 dx {\cal H}^q(x,t),\label{01}$,
 $F^q_2(t)=\int_0^1 dx {\cal E}^q(x,t).\label{02}$

 The proton and neutron Dirac form factors are defined as
 \begin{eqnarray}
 F_1^p(t)=e_u F_1^u(t)+e_d F_1^d(t),  \ \ \ \ F_1^n(t)=e_u F_1^d(t)+e_d F_1^u(t),
 \end{eqnarray}
  where $e_u=2/3$ and
 $e_d=-1/3$ are the relevant quark electric charges.
 As a result, the $t$-dependence of the $GPDs(x,\xi=0,t)$
can be determined from the analysis of the nucleon form factors
for  which experimental data exist in a wide region of momentum transfer.
 It is a unique situation as it unites  elastic and inelastic processes.

In the limit $t\rightarrow 0$, the functions $H^q(x,t)$ reduce to
 usual quark densities in the proton: $$ {\cal\
 H}^u(x,t=0)=u_v(x),\ \ \ {\cal H}^d(x,t=0)=d_v(x)$$ with the
 integrals $$\int_0^1 u_v(x)dx=2,\ \ \ \int_0^1 d_v(x)dx=1 $$
 normalized to the number of $u$ and $d$ valence quarks in the
 proton.

     \section{Nucleon quarks gravitational form factors and radii  }

%   Taking the  matrix
 %The energy energy-momentum  tensor $T_{\mu \nu}$
 %instead of  the electromagnetic current $J^{\mu}$
 %together with some model of
  %GPDs of quarks,
%  Taking the matrix elements of
 The energy-momentum tensor $T_{\mu\nu}$
    \cite{Pagels,Ji97,R04}  %PolyakovEMT}
    contains three gravitation form factors (GFF)
     $A^{Q,G}(t)$, $  B^{Q,G}(t)$, and $ C^{Q.G}(t)$.  We will scrutinize the first one which corresponds to
     the matter distribution in a nucleon.   This form factor contains quark and gluon contributions
        $A^{Q,G}(t)=A_{q}^{Q,G}(t)+A_{g}^{Q,G}(t) $.
% \ba
% \begin{eqnarray}
% \left\langle p^{\prime}|\hat{T}^{Q,G}_{\mu \nu} (0)|p\right\rangle &=& \bar{u}(p^{\prime}) \biggl[
% A^{Q,G}(t) \frac{\gamma_{mu}P_{\nu} }{2}   % \\
%     + B^{Q,G}(t)\frac{i\left(P_{\mu}\sigma_{\nu \rho} + P_{\nu} \sigma_{\mu \rho}\right) \Delta^{\rho} }{4 %     M_N}  \nonumber \\ \nonumber
% &+& \ C^{Q.G}(t) \frac{  \Delta_{\mu} \Delta_{\nu} - g_{\mu \nu} \Delta^{2} }{M_N}
% % \pm \ \bar{c}(t) g_{\mu \nu}
%\biggr ] {u}(p)
%\end{eqnarray}
%%  \ea
  One can obtain the gravitational form  factors of quarks which are related to the second  moments of GPDs.
 % one can obtain the gravitational form  factors of quarks which are related to the second, rather than the first moments %of GPDs \cite{Wang10}
%\ba
%\begin{eqnarray}
%\int^{1}_{-1}dx \ x [H_q(x,\Delta^2,\xi)  = A^{q}_{2,0}(\Delta^2)+(-2\xi)^2C^{q}_{2,0}(\Delta^2), \nonumber %\\
%\int^{1}_{-1}dx \ x[E_q(x,\Delta^2,\xi)=B^{q}_{2,0}(\Delta^2)-(-2\xi)^2C^{q}_{2,0}(\Delta^2).
%\end{eqnarray}
%\ea
 For $\xi=0$, one has
%\ba
 \begin{eqnarray}
\int^{1}_{0}dx \ x{\cal{H}}_q(x,t) = A_{q}(t); \,\ \int^{1}_{0}  dx \ x {\cal{E}}_q (x,t) = B_{q}(t).
\end{eqnarray}
%\ea
Note the different C-parity of gravitational form factors with respect to electromagnetic ones.

 %Our results for $A_{u+d}(t)$ are shown in Fig.1.
  Our GPDs with different PDFs lead
 to the same $t$ dependence of  $A_{u+d}(t)$.
 At $t=0$ these contributions equal $A(t=0) \approx 0.54$.
% and $A_d(t=0)=0.14$.
% The corresponding calculations for $B_q(t)$ are shown
%  in Figs. 14. In this case, we have the difference at $t=0$ and some difference in the $t$ %  dependence
%  already at small momentum transfer.
%     The PDFs {\bf O12a}  give the large values (upper curve in Fig.14)
%  and   PDFs {\bf GP8NNL} gave the lower values (low curve in Fig.14).  Others
%        concentrated in two clusters.
%  One gave  $B_{grav}(t=0) = -0.15$ (the PDFs {\bf JR8a, MRST09a, MRST09b, GJR07b}, and %  second gave
%  $B_{grav}(t=0) = -0.11$ the PDFs {\bf ABKM09, ABM12, KKT12A, MRST02}.
%  In our previous work \cite{ST-PRDGPD}, we obtained  $B_{grav}(t=0) = -0.05$ that
%      is close  to the zero value.
%%  At $t=0$ these contributions equal $B_u(t=0)=0.22$
%% and $B_d(t=0)=-0.27$. Hence their sum
% That is  a sort of compensation for the $u$ and $d$ quarks supporting
% the conjecture \cite{Teryaev-s3,Teryaev:2006fk} about the validity of the Equivalence % Principle
% separately for quarks and gluons.
 % :  $B_{u+d}(t=0)=-0.05$.
%In  the accuracy of our approximations this result coincides with zero.
%
%
%Note that nonperturbative analysis within the framework of the lattice OCD indicates
% that the net quark contribution to the anomalous gravimagnetic moment $B_{u+d}(0)$
% is close to zero  \cite{Gockler04,Hagler05}.
% Now, our results contradict this conclusion. Probably, it points out the important % contribution of the gluon part. \\
%
% \section{Nucleon gravitation  radii }
 % Using the obtained GPDs of the nucleon
 A good description of  various form factors and elastic scattering of  hadrons
 gives support for  our definition of the $t$ dependence of GPDs.
 Based on this determination of GPDs,
   let us calculate the gravitational radius of the nucleon using the integral representation of the form factor
  and make the numerical differentiation over $t$ as $t \rightarrow 0$. This method allows us
  to obtain a concrete form of the form factor by fitting the result of the integration of
   the GPDs over $x$.

   As a result, the gravitational radius is determined as
%   \ba
%\begin{eqnarray}
%<{r_{A}}^2> =  -\frac{6}{A(0)} \frac{dA(t)}{dt}|_{t=0}.
%\label{Rp}
%\end{eqnarray}
%\ea
%hence
 $<{r_{A}}^2> =  - 6/A(0)   \ \ dA(t)/dt|_{t=0}$ where
 %where
\begin{eqnarray}
  A(t) = \int_{0}^{1} x \ 3 (q_{u}(x) + q_{d}(x)) \ e^{-\alpha \ t f(x)} dx
\end{eqnarray}
and the radius  can be obtained by  a  numerical derivative. % will be
%   \ba
%\begin{eqnarray}
%<{r_{A}}^2> =  -\frac{6}{A(0)} \frac{A(t_{1})-A(t_{1}+\Delta t}{\Delta t}|;
%\label{Rp}
%\end{eqnarray}
%\ea
%   where $A(t) = \int_{0}^{1} x \ 3 (q_{u}(x) + q_{d}(x)) \ e^{-\alpha \ t f(x)} dx$

     The GPDs will be taken with   various  forms of the PDFs obtained by
    different Collaborations (see our paper \cite{GPD-PRD14}).
   To compare the $t$ dependence of the starting point of the differentiation of the results, we take 3 variants with different small intervals of $t$.
%  $$ 1) -t_{1} =1. 10^{-3} \ \ \ {\rm and} \ \ \ \Delta t_{1} =1. 10^{-3};$$
%  $$ 2)-t_{1} =1. 10^{-2}  \ \ \ {\rm and} \ \ \   \Delta t_{1} =1. 10^{-3}; $$
%  $$  3)-t_{1} =4. 10^{-2} \ \ \  {\rm and}  \ \ \  \Delta t_{1} =1. 10^{-3};  $$
   The results obtained for 9 PDFs by the different Collaborations  %are present in Table
   has the arithmetic mean $<r_{m}^{2}>^{1/2} = 0.524$ fm.
   It can be compared with  % $<r_{m}^{2}>^{1/2} =  0.55$ fm.
 the   result of D.E. Kharzeev  \cite{Kharzeev-21} %(PRD 104(5) (2021)
   $<r_{m}^{2}>^{1/2} =  0.55$ fm.
%     $<r_{m}^{2}>^{1/2} = {\bf 0.55 fm.}$
%        RPD (2020)
        and  with the electromagnetic radius \cite{Kharzeev-20}
         $ <r_{C}^{2}>^{1/2} =  0.8409$ fm.

   To examine the  dependence of the results on our model parametrization of GPDs,
     we  made the calculation for the simplest basic variant with minimum parameters
    and a more complicated variant with  4 additional parameters.
   It was shown in \cite{GPD-PRD14} that the final result of  fitting of GPDs
   on the basis of practically all experimental data on the electromagnetic form factors of the proton and
   neutron is weakly dependent on  adding  supplementary fitting parameters.
    %  but some others  had a heavy dependence.
%   We put the GPDs of the first class     in the upper rows of  Table 1.
%    It is clear from the comparison of the variants 1) and 2) (see Table 1)
%     that there is no  difference between
%    the results obtained at $-t=10^{-5}$ GeV$^2$ and $-t=10^{-3}$ GeV$^2$.
%    However, already at $-t=4. \ 10^{-2}$ GeV$^2$ the results have  significant %    differences.
%    Hence, if we make the fitting procedure to obtain the analytic form of the form %    factors,   the region of a very small momentum transfer has to be leading.

%   The comparizon of the "basic"  with "basic + 4 parameters" variants
%   shows that for PDFs of the various Collaborations leads to the different
%   results.
%   Note that the "basic"  and the "basic + 4 parameters" variants can lead
%   to the same different results for the PDFs of  different Collaborations.
%For example, the PDFs of (KKT12a \cite{KKT12}) give the radius which is practically %independent of
%   additional free parameters. Hence, this result can be considered as most stable and %   probable.
%   Note that this size of the radius practically coincides with the arithmetic means
%   of all radii (in the first column), which is equal to $\bar{(<{r_{A}}^2)>^{1/2}} =0.524$ %   {\it fm}.

\section{Gluon GPDs and gravitational radius}

  It is not that  simple to extract the gluon contribution to the
   gravitational radius
   from existing experimental data.
    As an example, in ref. \cite{Wang-Ch-21} the
   proton mass radius was obtained including   contributions of both quarks and gluons. They found
   $\sqrt{<R^{2}_{m}} =0.67\pm0.03$ fm  and the gravitational form factor was described by the dipole form.
  Really, the GFF can be presented in some multipole form
     \begin{eqnarray}
     A_{g}(t) = L^2/(L^2-t)^n.
  \end{eqnarray}
   It is  important to determine a separate gluon contribution to
  the gravitational form factor and determine the gluon gravitation radius.
  A separate analysis of quarks and gluon proton GFF was made in \cite{W-Z-23}.
  Modelling of both form factors by the dipole form and taking
  $A_{g}(0)=0.414$, they obtained $m_{g}=1.43\pm0.10$ amd $m_{q}=0.85\pm0.05$ GeV as free parameters.
   As a result, $\sqrt{<R^{2}_{m}} =0.69\pm0.04$ fm.
     There are two viewpoints on the size of $n$.
        First, in \cite{Sh-D-18} $n$ was taken equal to $2$ as a  dipole case;
        note that the generalized nucleon radii defined from the gluon GFFs are
        larger than those obtained from the corresponding quark quantities.
        The dipole  approximation was also used in \cite{Man0-Zah-21} for approximation
        of gluon GFF up to $-t =2$ GeV$^{2}$.
        Following the usual approximation, the dipole form was also used
        in \cite{Wang-Ch-21} and  simultaneously for both quark and gluon GFF in \cite{Wang-W2-22,W-Z-23}.
        Based on a lattice calculation in \cite{Guo-Ji-Liu-21}, the dipole form was used also.
         However, a large collaboration, which determined the proton's gluonic GFF,
         used the tripole approximations $n=3$ inspired by the latest lattice calculations \cite{Latt-PEF-21}. Following this viewpoint, to extract  gluonic GFF from the near-threshold $J/\Psi$ photoproduction measurements in \cite{Guo-Yang-23},  two GFFs $A_{g}(t)$ and
         $C_{g}(t)$ were parameterized in tripole form too.

  We used six different gluon parton distributions $xg(x)$.
  \cite{Broun2206.05672,BS-Bell,CTQE12,Vafaee} and two forms from \cite{Bonvinia}
%      u1(gr-sd)    u2(ld)     u3(bl-sol) u4(5)(l-red)  u5(6)(dt-vil) u6(7)(l-red)    %u7(9)(sd-blu)
%  They are presented in Fig. 2.
  Most of them have a polynomial form in $x$ while a more complicated form was obtained in \cite{CTQE12} taking $Y =\sqrt{x}$
    \begin{eqnarray}
%     \ \  \ \\ \nonumber
         F1&=&Sinh(3.032) (1 - Y)^3 + Sinh(-1.705) 3.Y (1 - Y)^2 +(3 + 1.062) Y^2 (1 - Y) + Y^3;   \nonumber
  \end{eqnarray}
   \begin{eqnarray}
                        xg(x)&=& 2.69 x^{0.531} (1 - x)^{3.148} F1 .
 \end{eqnarray}
   Note  especially the $xg(x)$ obtained in \cite{BS-Bell} which is  based on
%    the Bourrely-Soffer model    using
  the  parton statistical model and has  a non-standard form
   \begin{eqnarray}
 x g(x)= 15.853 \ x^{0.792}/(Exp(x/0.099)-1.).
 \end{eqnarray}
  As an example, we also used pion PDFs presented in work \cite{AriolaPi}.
  Further, we used the notation $u1,u2,u3,u4,u5,u6$ and $Pi$ corresponding the order of references.

    At first,  we used  complete sets of existing experimental data on
     electromagnetic form factors. The total number of data points is 912 and they include the region
      of a large momentum transfer (up to 30 GeV$^2$).
      Using the $t$ dependence of GPDs obtained for the quarks contributions,
      we get the corresponding gluon GFF.
 \begin{eqnarray}
     A(t) = \int_{0}^{1} x g(x) \ e^{-\alpha \ t f(x)} dx
  \end{eqnarray}
      We find that up to $|t| \leq 2$ their form is similar but al large $t$ they essentially diverge.  Possibly, it shows that our picture for gluon distributions is
      valid only for $|t| \leq 2$, which is a natural property of GPDs. To be consistent, we made a new fit for quark GPDs
      limiting the experimental data to the  region $|t| \leq 2$. We used the new parametrization
      of the obtained gluon gravitational form factors.
      We took two variants of the $t$ dependence of gluon GPDs. In the first variant $f(t)$,
      the exponents are taken the same as for the quark case,
      only with the size of the slope  equal to $0.4$   GeV$^{-2}$
      (which may be related to the glueball mass).

In the second variant, we simplify our picture and take the $t$-dependence in the simple Regge form $f(t)=-Log(x) t$ with the same slope $0.4$ GeV$^{-2}$.
      Using these results, we obtain that gluon gravitational form factors are described by
      the multipole form with the fitting parameters $L$ and $n$.
%   \begin{eqnarray}
%     A_{g}(t) = L^2/(L^2-t)^n.
%  \end{eqnarray}
%    The last variant is presented in Table 1.
    %    It is need note that
    In our analysis of all  $xg(x)$
   for both forms of $f(t)$, the size of  $n$ appears to be more close to $3$.
  %  and $L^2=0.9$.
     For the proton case, the arithmetical mean of the gluon gravitational radii in the first case is equal to $0.87 \pm 0.06$ fm.
    For the pion case $n=1$ and $L=1.4$ GeV, and the size of the radius equals $0.67$ fm.
    Note that the pion $R^{\pi}_{g} = 0.556$ fm was obtained in \cite{Hackett-2307.11707}.
    For the first variant of the $t$-dependence, we obtained similar results for $n$ and $L^2$
    and the arithmetical mean of the gluon gravitational radii  equals to $0.85\pm0.05$ fm.

\section{Conclusions}
   Our work supports that GPDs reflect the basic properties of the hadron structure
   and provide some bridge between  many different reactions.
   The determined new form of the momentum transfer dependence of GPDs
   allows one to obtain  different form factors, including Compton form factors,
   electromagnetic form factors, transition form factor, and gravitational form factors.
   The chosen form of the $t$-dependence of GPDs of the pion (the same as the $t$-dependence of the nucleon)
   allow us to describe the electromagnetic and GFF of pion and pion-nucleon scattering.

   Our investigation of the nucleon structure shows that the density of  matter
    is more concentrated than the charge density and,
    correspondingly the gravitational radius is smaller
     than the electromagnetic radius.
  This  naturally follows from the Regge dependence of GPDs being concentrated at low x,
 which was observed in our earlier work  \cite{GPD-ST-PRD09},
    The quark gravitational form factor $A_{q}(t)$ can be represented by dipole form
    with $L^2 = 1.6$ GeV$^2$.
    Using  various  gluon PDFs,
    we obtain the gravitation form factor $A_{g}(t)$, which can be represented
    by multipole form
    with power being more close $n=3$ and $L^2$ around $0.9$.
%      It is interesting note that in \cite{Burkert-23} the gravitational form factor $D(t)$,  encodes  the %mechanical properties of nucleon, represented in the multipole form
%      with $n=2.76 \pm 0.23$. And authors make the conclusion that it is consistent with a tripole behavior of %D(t).
     Our analysis shows that the gluon contribution to
    the hadron gravitation form factor is approximately  $44\%$ and the quark contribution is $54\%$.

   The different behaviour of gluonic GFF may result from the contribution of the quark-gluon wave function
    of the nucleon. In this case, an extra (in comparison to the quark case)
     gluonic propagator will provide a stronger t-dependence. This corresponds to the manifestation of quark (quark-gluon in this case) counting rules    \cite{Matveev:1972gbm,Brodsky:1974vy}

The gluon gravitational radius is
    approximately equal to the electromagnetic radius.
  In particular, the arithmetical mean of gluon and quark proton gravitational form factors coincides with the results obtained in \cite{Wang-Ch-21}.
 %    of \cite{RWang21} of proton gravitation form factors

The reason for such a behaviour ("quark-gluon duality") may be that the region ($x \sim 0.2 - 0.3$) corresponding to the  maximum of valence quark distributions marks also the transition (when going to lower x) to the  dominance of gluons over quarks.
 The various experimental, theoretical and model tests of this picture are highly desirable.

%\begin{acknowledgments}
% To all
%\end{acknowledgments}

%\appendix

%\section{Methodic}

%\section{Calculations}

\end{document}